\documentclass[a4paper,12pt]{article}
\usepackage{amsmath,amssymb,amsthm}                 \usepackage{epsfig}
\usepackage{subfigure} \usepackage{graphicx}

\begin{document}

\title{On the Twin Paradox in a Universe with a Compact Dimension}
\author{Dhruv
Bansal,  John  Laing,   Aravindhan  Sriharan}
\maketitle

\begin{abstract}
We consider the twin paradox  of special relativity in a universe with
a  compact  spatial  dimension.   Such  topology  allows  two  twin
observers to remain inertial yet meet periodically.  The paradox is  resolved  by
considering  the  relationship  of  each twin  to  a  preferred
inertial  reference frame which exists in such a universe because global Lorentz invariance is broken.  
The twins can perform ``global'' experiments to determine their velocities with respect to the preferred reference frame (by sending light signals around the cylinder, for instance). Here we discuss the possibility of doing so with local experiments. Since one spatial dimension is compact, the electrostatic field of a point charge deviates from $1/r^2$. We show that although the functional form of the force law is the same for all inertial observers, as required by local Lorentz invariance, the deviation from $1/r^2$ is observer-dependent. In particular, the preferred observer measures the largest field strength for fixed distance from the charge.
\end{abstract}

\section{Introduction}
In the classic presentation of the twin paradox, \cite{SpacetimePhysics},
two  observers each witness the other receding
at constant  velocity and  returning at the  same velocity at  a later
time.   Each  observer will  claim  he  was  stationary and,  by  time
dilation,  that the  \emph{other} should  be younger  upon meeting.
The resolution is that one observer turned
around at  some point during  the journey and,
consequently, was  not inertial for  the entire duration of  the trip.
This kinematic asymmetry allows  both twins to unambiguously determine
which  of them aged  more during  the journey:  the twin  who remained
inertial throughout.

In a space-time with  one spatial dimension  compactified,  $S^1 \times  \mathbb{R}^{2,1}$,  this
kinematic solution  no longer works.   Both twins can  remain inertial
for  the \emph{entire}  journey if  they confine  their motion  to the
compact  dimension  (see Fig.  \ref{figure:cylinderplane}).  In this case, 
the resolution  lies in recognizing that  compactifying a spatial dimension breaks global Lorentz invariance, \cite{BransStewart}.  
In particular, there is now a \emph{preferred} inertial reference frame, 
\cite{BransStewart,LevinParadox,Periodic}, namely that for which the circle 
is purely spatial ({\it i.e.,} the observer whose worldline does not wind around the circle, \cite{Topology}). The
relationship  of each  observer to  this reference
frame establishes  the asymmetry required  to resolve the
paradox:  the  observer in the  preferred frame is essentially at
rest with  respect to the universe and ages more than the moving observer during the journey.

It is well known that observers can determine whether or not they are in the preferred rest frame by sending light beams in opposite directions along the compact dimension, \cite{BransStewart,Periodic}. After  waiting for  the light  beams  to traverse  the entire  compact
dimension, only the observer in the preferred frame will receive both signals simultaneously. Moreover, the time interval between the two signals is related to the velocity of the observer relative to the preferred frame.

Such a global experiment is of little practical use if the size of the circle is on the order of cosmological scales since an observer would have to wait about a Hubble time before receiving his signals.
Here we present a \emph{local} experiment that either twin can perform to determine his relationship to the preferred frame based on measuring deviations from the $1/r^2$ force law. 
The electric (or gravitational) field in a universe with a compact dimension is not exactly $1/r^2$ but depends on the size $L$ of the compact dimension because field lines are confined in this direction.  Since local Lorentz invariance still holds, the functional form of the field is the same for all inertial observers, but the parameters which appear in the force law, which can be thought of as effective fine-structure (or Newton's) constants, do depend on the observer. This can be understood qualitatively because the size of the compact dimension is not invariant under boosts. A boosted observer sees a larger effective circle (segment, actually) and thus a weaker field. Conversely, an observer in the preferred frame measures the strongest field at fixed distance from the source. Hence by making measurements of the electric field of a point charge stationary in their frame, observers may determine the effective length of the compact dimension in their frame, $L_{eff}$.  Comparing $L_{eff}$ with $L$, the length of the compact dimension in the preferred frame, precisely specifies the relationship of the observer to the preferred frame and resolves the paradox: the boosted observer ages less during the paradox by a factor $\gamma = L_{eff}/L$.

\section{Analysis of the Paradox and a Global Resolution}
\subsection{Description of the Space-time}\label{DescriptionOfSpacetime}
The manifold we  are considering in this problem  is the cylinder, $S^1
\times \mathbb{R}^{2,1}$, with the Minkowski  metric, $ds^2 = -dt^2 + dx^2
+ dy^2 + dz^2$.  It can be constructed from $\mathbb{R}^{2,1}$ by imposing
the equivalence relation
\begin{equation}\label{equivrel}
(t,x,y,z) \sim (t,x+nL,y,z)
\end{equation}
where $L$ is  the circumference of the compact  dimension and $n$ is an integer.  Each
equivalence class of points $\left[ (t,x+nL,y,z)\right]$ in $\mathbb{R}^{3,1}$ is represented
by a single point $(t,x,y,z)$ on the cylinder, chosen such that $0\le x < L$.

\begin{figure}[htbp]
\centering \subfigure[]{
	\label{figure:cylinder}
\includegraphics[width=1in,height=2.5in]{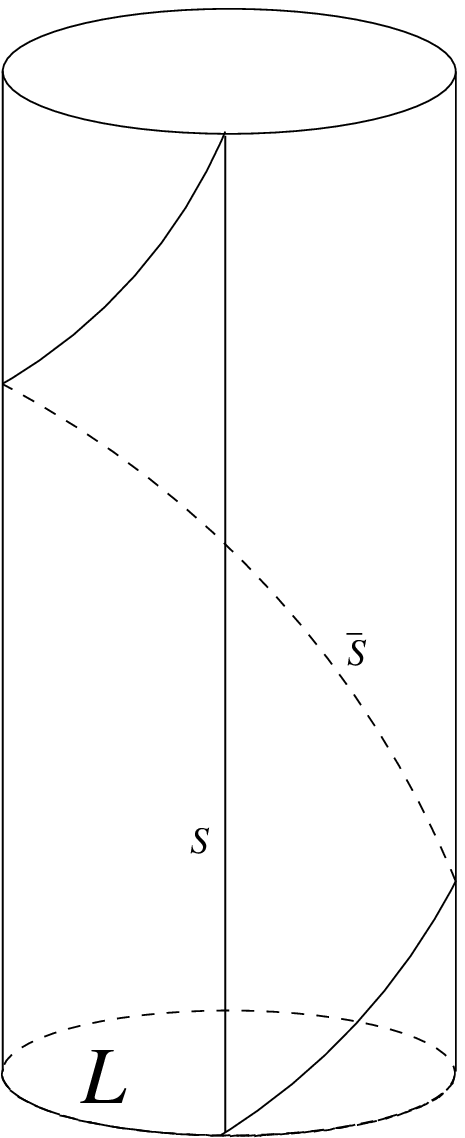}}
\hspace{.3in} \subfigure[]{
	\label{figure:plane}
\includegraphics[width=3in,height=2.5in]{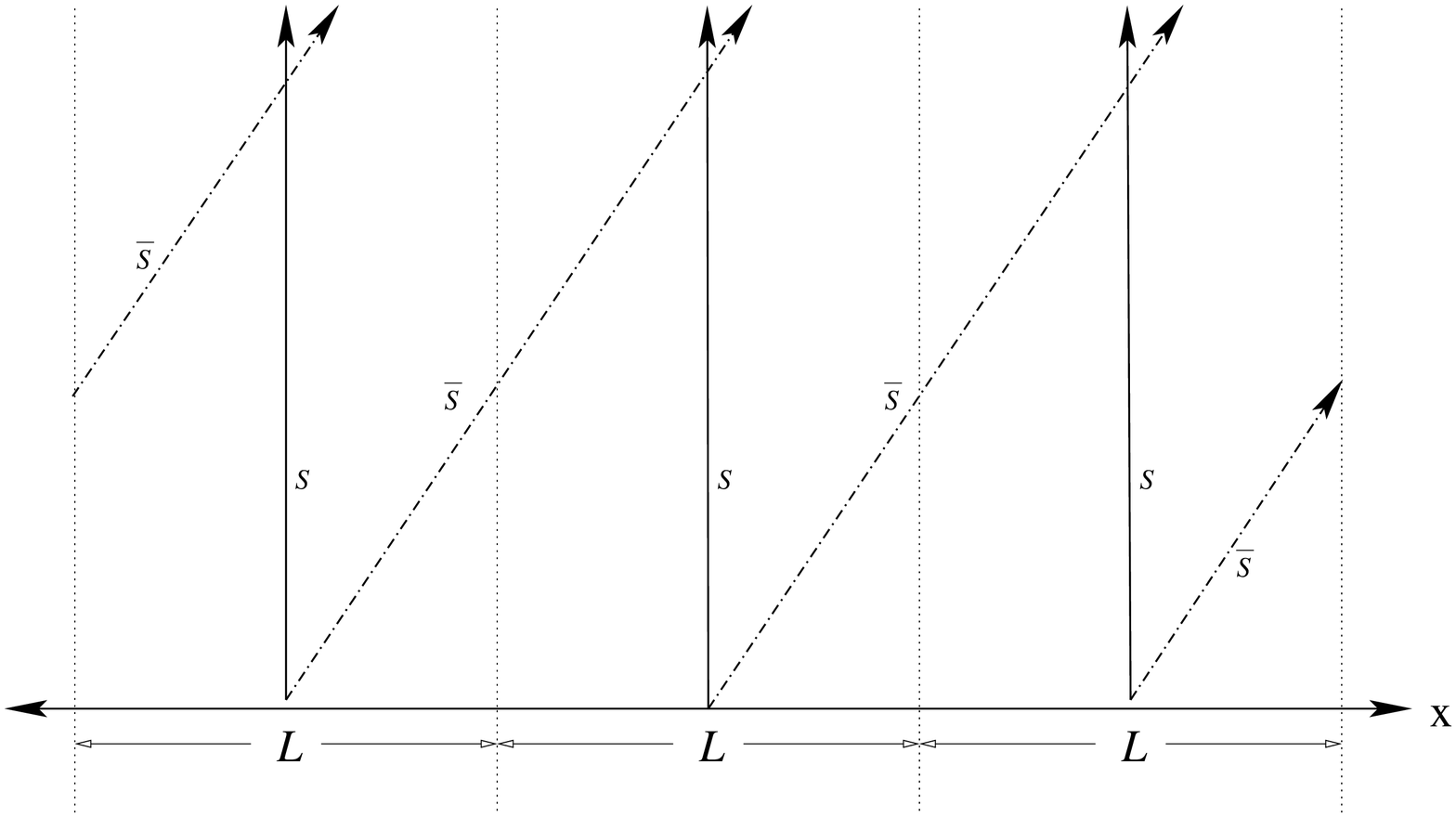}}
\caption{The  worldlines  of  the  preferred observer  ($S$)  and  the
non-preferred   observer  ($\overline{S}$)   in   the wrapped picture (a) and the unwrapped picture (b).}
\label{figure:cylinderplane}
\end{figure}

We thus have  two equivalent  pictures  of the  manifold $S^1  \times
\mathbb{R}^{2,1}$:  the ``wrapped'' picture, Fig. \ref{figure:cylinder}, where each point is a unique event, and the ``unwrapped'' picture in the covering space, Fig. \ref{figure:plane}, where an infinity of points represents the same event.  We can consider the latter picture as an infinite sheet of paper which
we ``wrap'' into a cylinder to construct the former picture.  It will prove useful to be able to switch back and forth between these two pictures.

Lorentz invariance is broken globally since one dimension is compact, which leads to
the existence of a preferred rest  frame. To see this, consider that the  equivalence relation  \eqref{equivrel} is
manifestly dependent  on coordinates and so  is \emph{itself} defined
in  a  particular  frame,  call  it  $S$.   In  this  frame,  a  point
$p=(t_0,x_0,y_0,z_0)$  in  the   wrapped  picture  corresponds  to  an
infinity  of  points $p_n  = (t_0,x_0+nL,y_0,z_0)$  in the  unwrapped
picture.   These  points are  all  simultaneous  in  $S$ and they differ from each other only by spatial translations.  What  about
$\overline{S}$, a  frame moving with  respect to $S$ at  some velocity
$\beta=v/c$?   In  this  frame,  the   point  $p$  has  coordinates  $p  =
(\overline{t_0},\overline{x_0},\overline{y_0},\overline{z_0})=(\gamma
t_0 - \beta \gamma x_0,\gamma x_0  - \beta \gamma t_0, y_0,z_0)$, where $\gamma=(1-\beta^2)^{1/2}$.  In the unwrapped  picture, this point
corresponds to  the points  $p_n =  (\gamma t_0 -  \beta \gamma  x_0 -
\beta  \gamma   n  L,\gamma   x_0  +  \gamma   n  L  -   \beta  \gamma
t_0,y_0,z_0)=(\overline{t_0}  -  \beta  \gamma  n  L,\overline{x_0}  +
\gamma  n  L,  y_0,z_0)$.  We see that the equivalence relation \eqref{equivrel} in an
arbitrary inertial frame $\overline{S}$ becomes
\begin{equation}\label{equivrelgen}
(\overline{t},\overline{x},\overline{y},\overline{z}) \sim
(\overline{t}-n \beta\gamma L,\overline{x}+n \gamma L,\overline{y},\overline{z})
\end{equation}
Thus, the image points simultaneous in frame $\overline{S}$ are not only translated through space, but through time as well.  This is a recognition of the fact that lines of equal time for observers with $\beta \neq 0$ do not close on themselves but spiral around the cylinder.  There is only one observer, characterized by $\beta = 0$, whose line of equal time closes on itself, and for whom the identification \eqref{equivrelgen} is a purely spatial one.  We refer to the frame of this observer as the \emph{preferred rest frame}.

It should be noted that the effective size of the compact dimension in a frame $\overline{S}$ is $\gamma L$, as can be seen directly from \eqref{equivrelgen}.  We thus define the effective length of the compact dimension:
\begin{equation}
L_{eff}\equiv\gamma L
\end{equation}
The preferred observer measures the smallest value of $L_{eff}$, namely $L$.

\subsection{Minkowski Diagrams and Transition Functions}
The essential problem with the twin paradox in this space-time is that both 
twins can draw Minkowski diagrams which depict the other twin winding 
around the circle and coming back. Hence each twin will predict that 
the other is younger.  We can resolve this contradiction by noting that, for the preferred observer, the images of the fundamental domain $(0,L)$ are simply translated spatially by \eqref{equivrelgen}.  For a non-preferred observer, however, these images are also translated in time.  This implies that a diagram like Fig.\ref{figure:plane} is not valid in a non-preferred frame, and that a non-preferred observer cannot naively draw Minkowski diagrams.  Instead, the non-preferred observer must take into account certain transition functions.

Because one dimension is compact, observers in our 
space-time have a problem with multi-valued coordinates.  In Sec.~\ref{DescriptionOfSpacetime} we glossed over this point and implicitly treated all
lengths in the $x$-dimension modulo $L$.  To be more precise, we should 
really cover the manifold $S^1 \times \mathbb{R}^{2,1}$ with two 
single-valued coordinate patches and glue them together with 
appropriate transition functions.

In some coordinate system $S$, let patch $A$ cover the entirety of the $t$, 
$y$, and $z$ dimensions and cover an open interval $(0,L_{eff})$ of the 
compact $x$-dimension. Likewise, let patch $B$  cover the entirety of the uncompact dimensions and cover an interval $(-\epsilon,\epsilon)$ in the $x$-dimension.  As an analogy, one may think of patch $A$ as a piece of paper wrapped around the cylinder and patch $B$ as a strip of tape applied on the seam of patch $A$.

\begin{figure}[htbp]
\centering
\includegraphics[width=1in,height=3in]{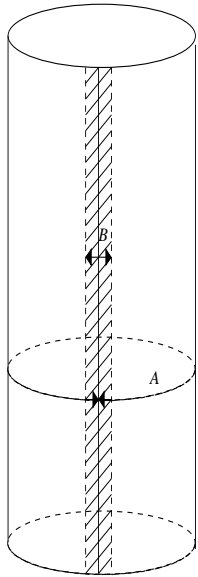}
\caption{We require two patches, $A$ and $B$, to cover the manifold $S^1 \times \mathbb{R}^{2,1}$}
\end{figure}

As we pass from patch $A$ to patch $B$, we wind around the cylinder or, 
equivalently, move to another image patch in the unwrapped picture, Fig. \ref{figure:plane}.  The index $n$ can thus be thought of as winding number, \cite{Topology}.  If observers keep track of the winding number of light signals, etc., Eq. \eqref{equivrelgen} describes how to relabel paths as their winding number changes.  Since a change in winding number corresponds to leaving patch $A$, crossing through patch $B$, and returning to patch $A$, Eq.~\eqref{equivrelgen} is recognized to be exactly the transition function we need:
\begin{equation}\label{TransitionFunctions}
f_{\pm}(t,x,y,z)=(t \pm \beta \gamma L,x \mp \gamma L,y,z)\,,
\end{equation}
where $f_{+}$ and $f_{-}$ are the transition functions used when winding around in the positive and negative $x$-directions, respectively.

\begin{figure}[htbp]
\centering \subfigure[]{
	\label{figure:twins1}
\includegraphics[width=2in,height=3in]{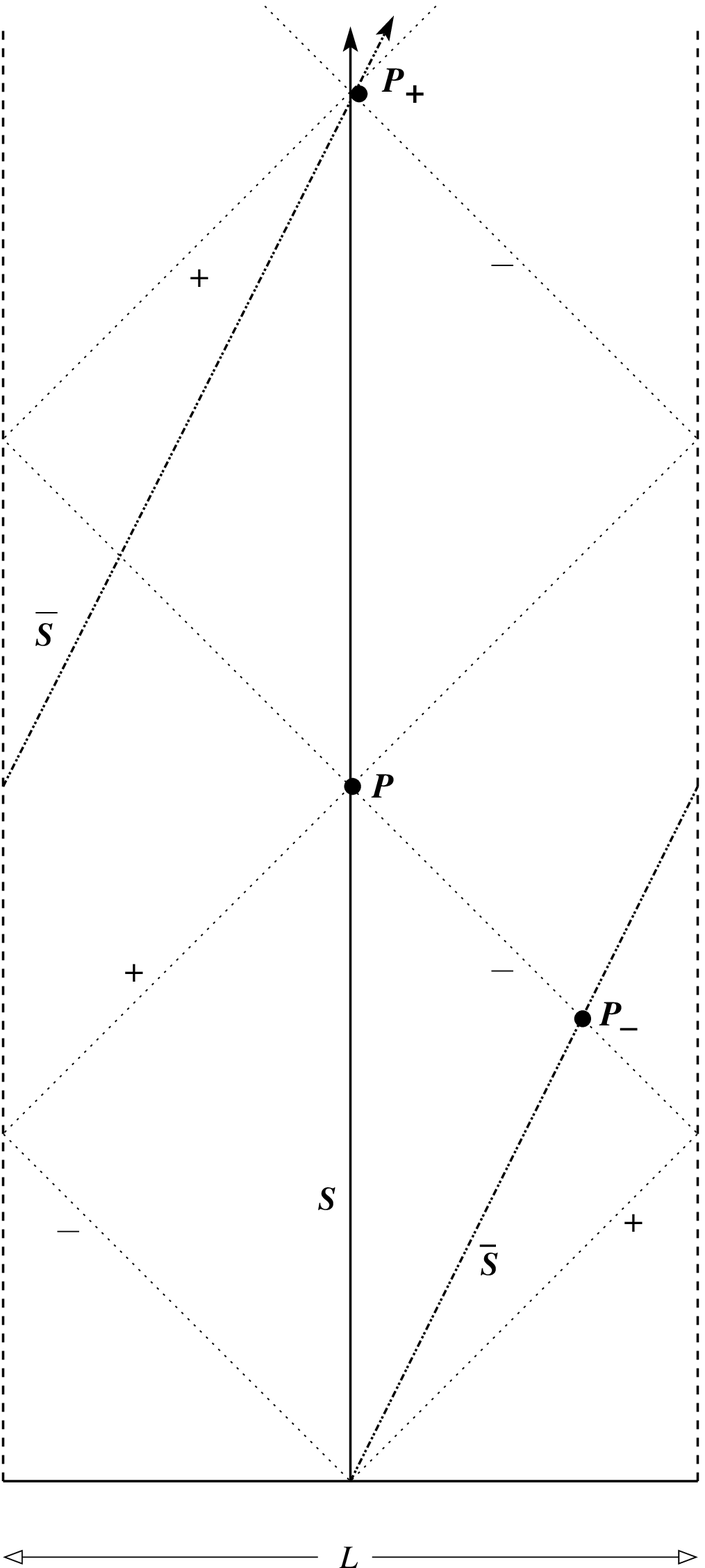}}
\hspace{.3in} \subfigure[]{
	\label{figure:twins2}
\includegraphics[width=2in,height=3in]{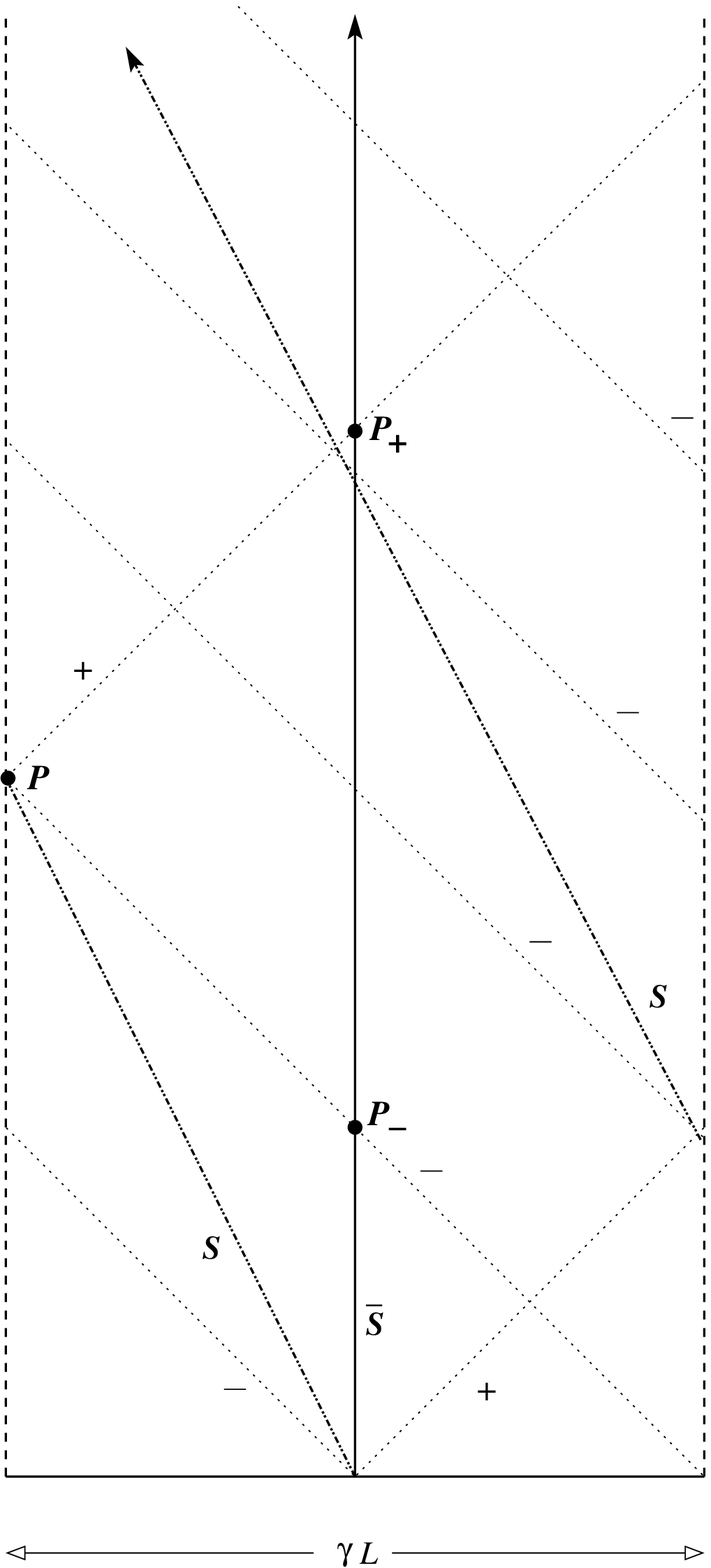}}
\caption{Minkowski diagrams depicting the twin paradox from the preferred 
frame (a) and a non-preferred frame moving with $\beta=0.5$ (b).  The 
thick lines represent the worldlines of the twins and the dotted lines, 
labeled by $+$ and $-$, represent two light signals which wrap around the 
circle in the positive and negative $x$-directions, respectively.  Note that the preferred observer ($S$) receives both $+$ and $-$ 
light signals at the same point ($P$) while the non-preferred observer 
($\overline{S}$) receives them at two different points ($P_+$ and 
$P_-$).} \label{figure:minkowskis} 
\end{figure}

When a given observer attempts to describe the physics within a 
single patch, say patch $A$, he must keep track of how to adjust the 
coordinates he assigns to objects as they exit and then re-enter the patch.  
For observer $S$, in the preferred frame, $\beta=0$, and there 
is no translation in time as objects wind around the universe.  This is 
why he can naively draw diagrams like Fig. \ref{figure:plane}.  For observer  $\overline{S}$, however, $\beta \neq 0$, and the transition functions involve translations in time.  In this frame, 
a diagram like Fig. \ref{figure:plane} would simply be 
incorrect.  Figure~\ref{figure:minkowskis} properly depicts the situation in both frames using the appropriate transition functions. No contradiction ensues.

If both twins know their velocity with respect to the preferred frame, 
then, by Eq. \eqref{TransitionFunctions}, they can find their transition 
functions and use them to draw correct diagrams. By using the transition functions, both observers in the twin paradox come to the conclusion that the twin in the non-preferred frame ages less 
during the journey by a factor of $\gamma$.

\subsection{Einstein Synchronization on the Cylinder}
We have thus far used two coordinate patches for the purely mathematical 
reason of avoiding multi-valued coordinates.  The need for multiple 
patches can, of course, be understood from a physical point of view by 
considering the  synchronization of clocks in this space-time.

The usual method for synchronizing clocks is Einstein synchronization: if 
an observer is midway between two clocks and receives light signals from 
each clock with the same reading simultaneously then the two clocks are 
said to be Einstein synchronized.  Usually, Einstein synchronization is a 
transitive process: if clock $A$ is synchronized with clock $B$, and clock 
$B$ is synchronized with clock $C$, then clock $A$ is synchronized with 
clock $C$.

Einstein synchronization immediately fails on a compact dimension because 
there are two midpoints between any pair of clocks.  We can circumvent this 
problem by choosing a ``left-most'' and a ``right-most'' clock.  
These clocks will demarcate the edges of what will become a coordinate 
patch.  We can synchronize clocks by using the midpoint in between these 
two boundary clocks, the midpoint in the coordinate patch we are 
constructing.  Transitivity is preserved because we confine 
all our procedures to this single patch which, without global data, is 
indistinguishable from an uncompact space.

A problem occurs when we let the left-most and right-most clocks approach 
each other, letting our coordinate patch encircle the entire compact 
dimension.  As soon as they overlap, that is, as soon as the left-most 
clock and the right-most clock are the \emph{same} clock, we will have 
constructed a global rest frame and, in a non-preferred frame, this clock will have to read one time to be synchronized 
with the clock on its left and another time to be synchronized with the 
clock on its right.  This is easily seen by considering lines of equal time in the wrapped picture.  For the preferred observer, such lines close on themselves and form circles.  For a non-preferred observer, however, they do not close but instead spiral endlessly around the cylinder.  The non-preferred observer's coordinate system corresponds to a segment of such a spiral.  If this is to span the cylinder, then the segment of the spiral must also span the cylinder. However, if we require each clock to only read one time, this implies that it must be discontinuous at a point.  The transition functions \eqref{TransitionFunctions} are a reflection of this fact. Therefore, while it is possible for the boosted observer to synchronize clocks in this way, evidently this comes at the expense of homogeneity. Indeed, it introduces a special line on the cylinder where time jumps.

In fact, there is a perspicuous analogy between the use of transition functions and patches on this space-time and a more familiar phenomenon: the time zones on the Earth.  Imagine a person standing on the equator keeping time by the Sun.  In his reference frame, fixed at a point on the Earth's surface, the Sun revolves about the Earth once per day.  He attempts to label points on the equator with their distance from him and with a particular time based on the position of the Sun as seen from that point.  At a particular moment, let him declare that it is high noon at his own position.  Points on the equator east of him will be assigned later times, while points west will be assigned earlier times.  As long as his reference frame is local and doesn't span the equator, nothing goes wrong in his scheme.  As soon as it does, however, the point diametrically opposite him on the equator demands to be labeled by \emph{two} points in time, one to coincide with the points immediately east of it and another for the points immediately west.  His solution is to draw an international date line through that point -- a transition function or discontinuity in his coordinate system.

\subsection{Global Experiment to Distinguish Twin Observers}

To determine his velocity with respect to the preferred frame, an observer can send out 
light signals in opposite directions along the compact 
$x$-dimension, \cite{BransStewart}.  
From Fig. \ref{figure:minkowskis} it is clear that the preferred observer, 
whose transition function does not involve translations in time, will 
receive the signals at the same time (at the event labeled by $P$). A non-preferred observer, 
however, will measure a time-delay in the reception of the two signals (the events $P_+$ and $P_-$).  
A simple calculation yields
\begin{equation}\label{RatioToBeta}
\beta = \frac{\tau(P_+) - \tau(P_-)}{\tau(P_+) + \tau(P_-)}\,,
\end{equation}
where $\tau(P)$ is the proper time at which event $P$ occurs. This expression can 
be used to determine the velocity with 
respect to the preferred frame. For the preferred observer, one has $\beta=0$ and indeed
$\tau(P_+) = \tau(P_-)$.
Once an observer knows his velocity 
with respect to the preferred frame, he can easily calculate the 
transition functions \eqref{TransitionFunctions} and draw appropriate 
Minkowski diagrams.

\section{Electromagnetism and a Local Resolution}
The experiment described above would take a prohibitively long time in
a universe of  any realistic size, as light signals have to
encircle  the  \emph{entire}  compact  dimension!   Furthermore,  this
global solution does  not seem as satisfying as  the local solution to
the  twin  paradox  in  standard  space-time  $\mathbb{R}^{3,1}$.   In  the
standard  space-time,   each  observer  may  easily   conduct  local
experiments to determine  whether or not he is  the accelerated twin --
he could hang a pendulum, for example, and watch for any deviations in
its path during the journey.

It seems that any
local  kinematic experiment  would not  serve to  resolve  the paradox
because there  are no  \emph{local} kinematic differences  between the
two  observers which  might be
exploited to distinguish them.   The global solution already presented
works precisely  because it is global  - the light  beams traverse the
entire compact  dimension, cross between coordinate patches, and thus force the observers to use transition functions, which encode the relationship between
the observer  and the  preferred rest frame.   
Here we propose to exploit the local consequences of  global phenomena
such as electric or gravitational fields.  A field permeates all of space and thus ``knows'' about the
global topology.  This global knowledge can be extracted by making 
measurements of the field at a few points.

\subsection{Electromagnetism on $S^1 \times \mathbb{R}^{2,1}$}
Consider the  electromagnetic field of a  point charge $q$  at rest at
the origin of  the preferred rest frame. One expects that the formula for
the electromagnetic field of this point charge should deviate from the usual $1/r^2$ since
the field lines cannot spread as much in the compact direction. Moreover, such deviations should
depend on the size of the circle, $L$.

\begin{figure}[htbp]\label{figure:ChargeBoost}
\centering
\subfigure[]{\includegraphics[width=4in,height=1.5in]{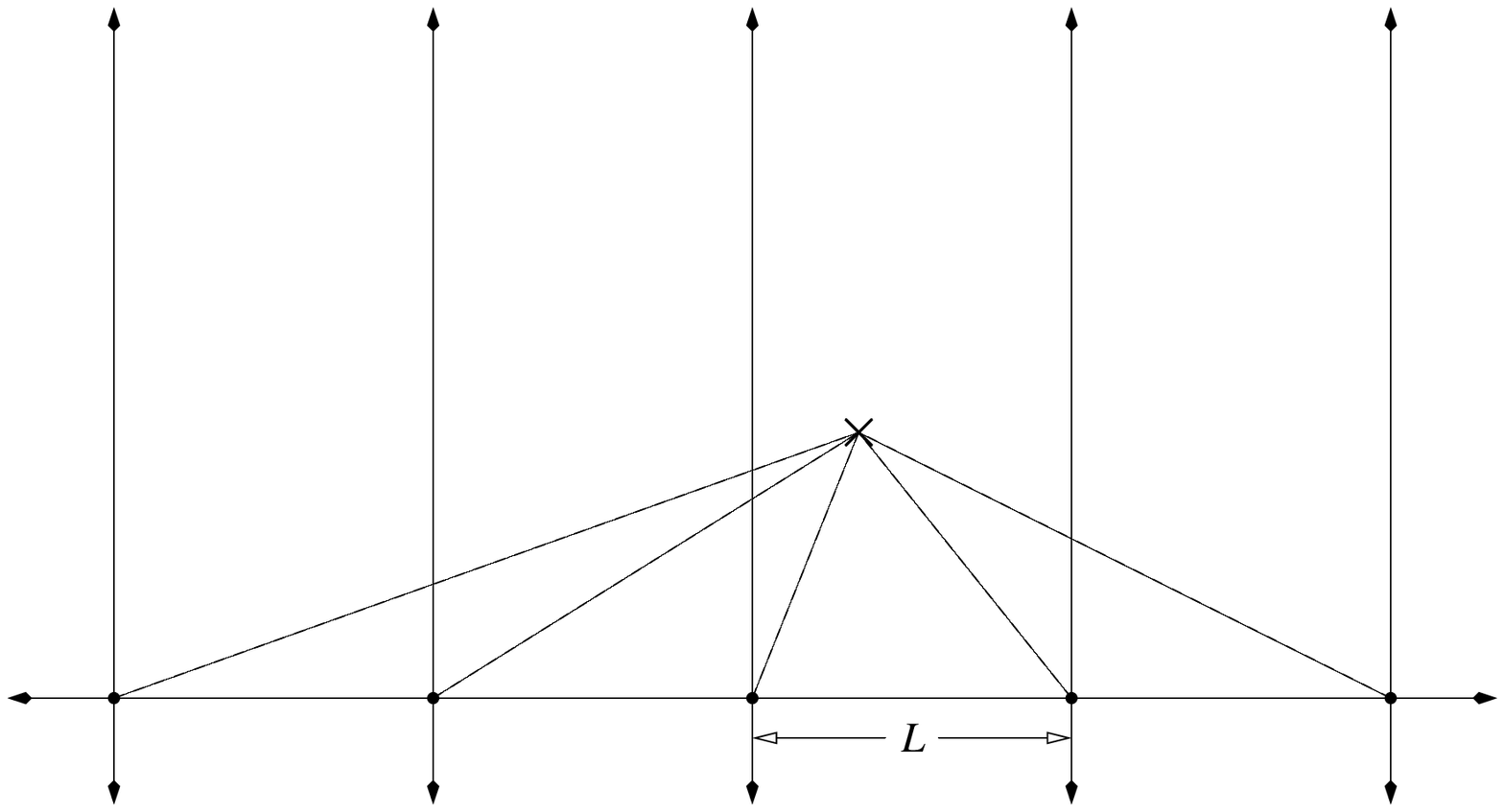}}
\subfigure[]{\includegraphics[width=4in,height=1.5in]{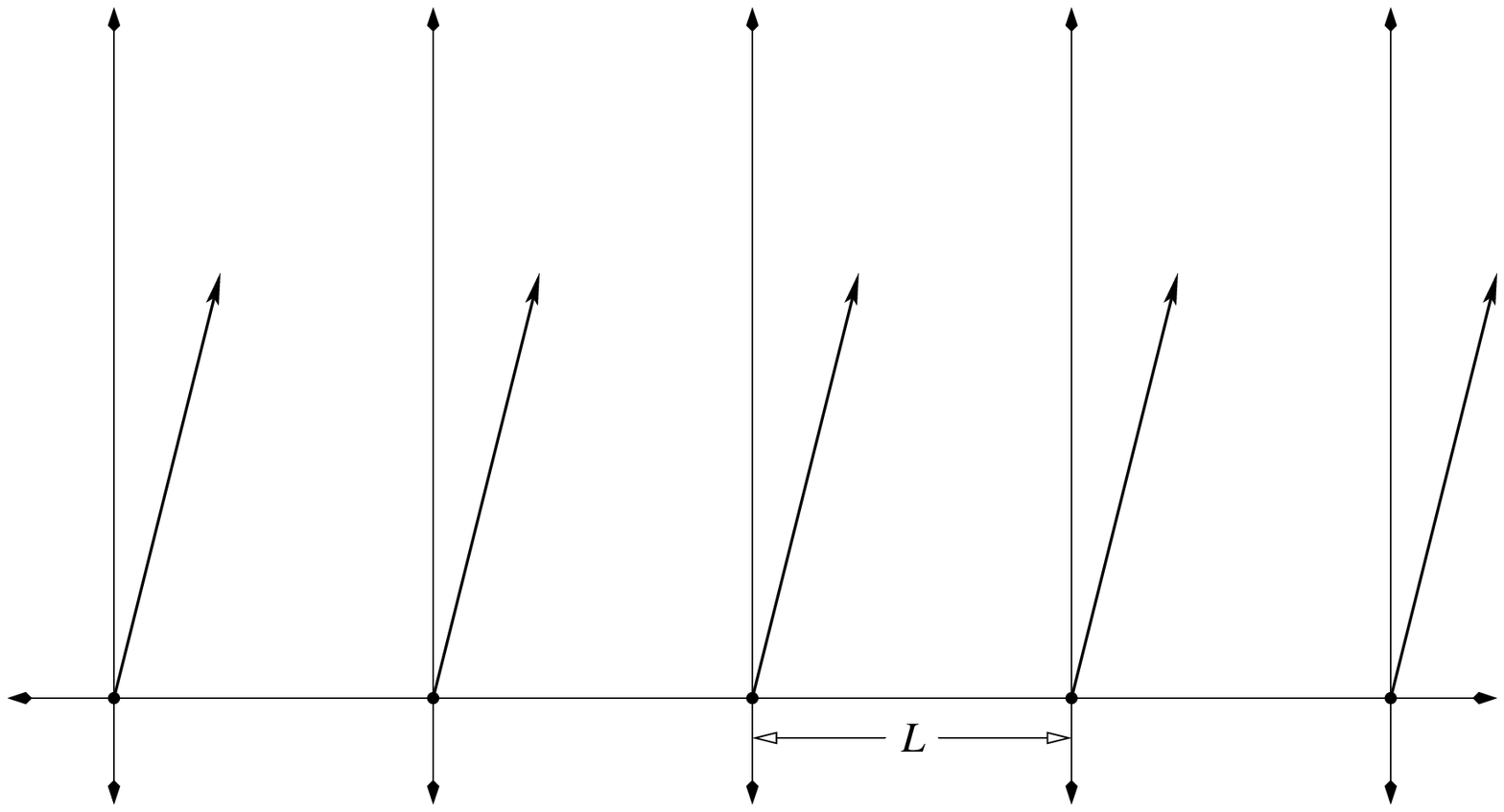}}
\subfigure[]{\includegraphics[width=5.14in,height=1.5in]{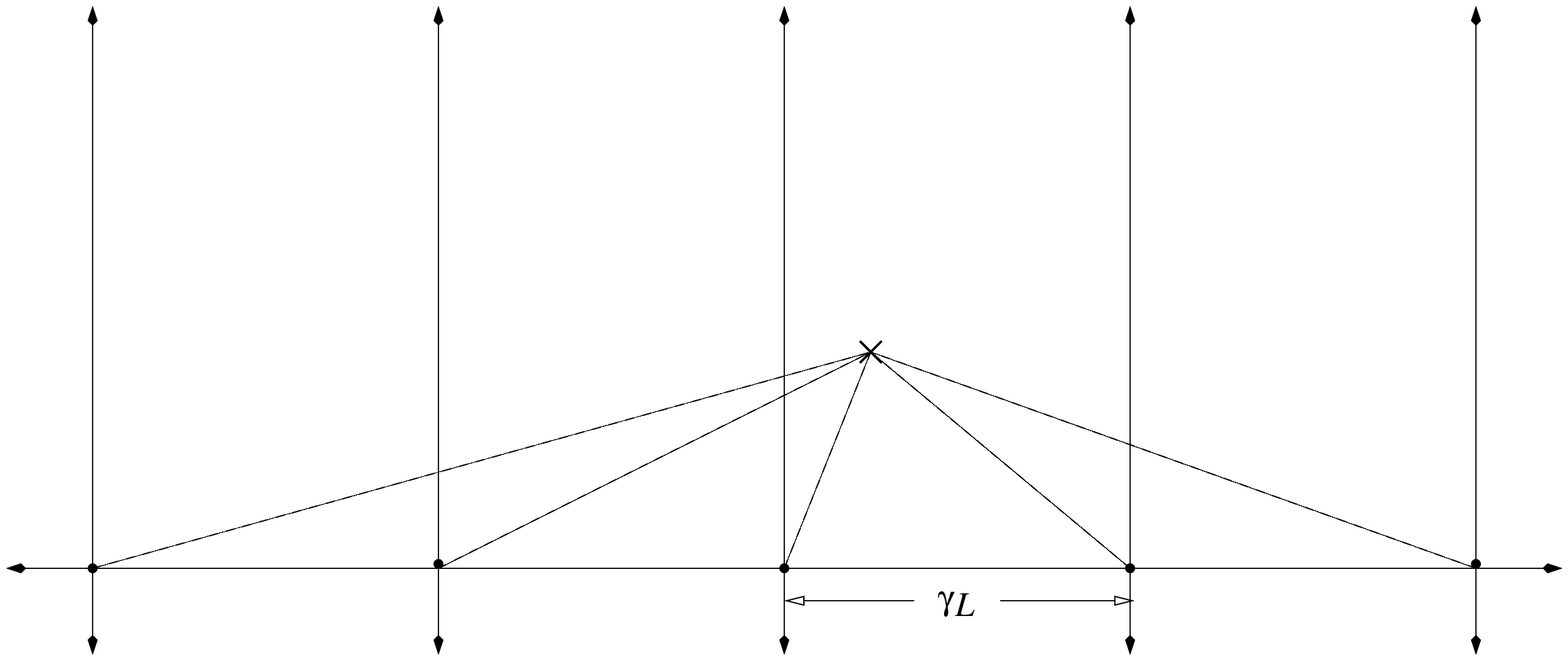}}
\caption{A single image charge stationary at the origin of the preferred frame may be considered as an infinite number of image charges in the unwrapped picture (a).  A charge moving at constant velocity in the preferred frame (b) may be considered as a charge stationary at the origin of a boosted, non-preferred frame.  In this frame, we may again consider the single charge as an infinite number of image charges (c).}
\end{figure}

To calculate the field, it  is easiest to work in the unwrapped picture and  consider each  image  charge as  a source for  the
electromagnetic field at  the field point (see Fig. 4).  There is no magnetic
field,  of course,  since the  point charge and  hence all its  image
charges are at rest in this frame.  We find
\begin{equation} \label{EFieldPRFSeries}
\vec{E}_S(x,y,z)=\frac{q}{4   \pi   \epsilon_0}\sum_{n=-\infty}^{\infty}
\frac{(x+nL)\hat{x}  + y\hat{y}  + z\hat{z}  }{\left[(x+nL)^2 +  y^2 +
z^2\right]^{\frac{3}{2}}}\,,
\end{equation}
which depends on $L$, as expected. It is easy to see that one recovers the usual Coulomb law in the limit $L \to \infty$.

What  about the  electric field  of a  point charge  stationary  in a \emph{non-preferred} frame?  Because Lorentz invariance is locally valid in this space-time, the field measured by a non-preferred observer should have the same functional form as Eq.~\eqref{EFieldPRFSeries} -- it can only differ in the values of some parameters. The only parameter to be found in Eq.~\eqref{EFieldPRFSeries} is the length of the compact dimension, $L$.  Thus we expect $L$ to be replaced with $L_{eff}$, the effective length of the compact dimension as measured by an observer in a non-preferred frame.

This answer is most easily obtained by noting that a point charge stationary in a non-preferred frame is of course moving at some constant velocity with respect to the preferred observer. From the preferred frame, we can boost directly into the rest frame of the charge and find that we have reproduced the situation we started with prior to deriving Eq. \eqref{EFieldPRFSeries}: a stationary point charge and an infinite series of image charges, each separated by the effective length of the compact dimension in that frame, $L_{eff}$. This is illustrated in Fig. 4.  Thus, we have
\begin{equation} \label{EFieldSeries}
\vec{E}_{\overline{S}}(x,y,z)=\frac{q}{4   \pi   
\epsilon_0}\sum_{n=-\infty}^{\infty}
\frac{(x+n L_{eff})\hat{x}  + y\hat{y}  + z\hat{z}  }{\left[(x+n L_{eff})^2 +  y^2 +
z^2\right]^{\frac{3}{2}}}
\end{equation}
for an arbitrary frame $\overline{S}$.  The only change from Eq. \eqref{EFieldPRFSeries} is a substitution $L\to L_{eff}$.  The field measured by any observer in this universe thus 
has a dependence on the parameter $L_{eff}$, the effective length of the 
universe in the frame of the observer.

A local experiment immediately suggests itself.  If we presume observers in this space-time know the value of $L$, then measuring the electric field of a stationary point charge at a few points is enough to determine $L_{eff}$, from which one can determine $\beta$ and resolve the twin paradox.

Restricting our attention to points on the $x$-axis, the infinite sum in Eq.~\eqref{EFieldSeries} can be
written in closed form using residue theorems and then expanded in powers of $x/L^2$:
\begin{eqnarray} \label{EFieldAxis} \nonumber
\vec{E}_{\overline{S}}(x,0,0)&=&\frac{q}{4 \pi \epsilon_0}\sum_{n=-\infty}^{\infty}\frac{\hat{x}}{(x+n L_{eff})^2}
\\ \nonumber
&=&\frac{q}{4 \pi \epsilon_0}\frac{\pi^2}{L_{eff}^2}\csc^2\left(\frac{\pi x}{L_{eff}}\right)\hat{x} \\
&=& \frac{q}{4 \pi \epsilon_0} \left[ \frac{1}{x^2}+ \frac{\pi^2}{3 L_{eff}^2}+\mathcal{O} \left(\frac{x^2}{L_{eff}^4} \right) \right] \hat{x}\,.
\end{eqnarray}
It is intriguing to note that the first order correction to the electric field (along the $x$-axis) in this topology is constant, with the fractional difference from the usual Coulomb field given by
\begin{equation}
\frac{\Delta E}{E} \approx \frac{\pi^2}{3}\left(\frac{x}{L}\right)^2.
\end{equation}
As expected, the difference increases with decreasing $L$. 

If this experiment is to be practical, however, then the ratio $\Delta E/E$ must not be vanishingly small.
The smallest allowed $L$ is $L=24$~Gpc from cosmic microwave background 
analysis, \cite{NoCircles} (though this figure may require revision, 
see, \cite{LevinCMB}). Unfortunately, for any realistic $x$, this ratio is 
unmeasurably small. Moreover, it is easily seen that the difference in magnitude between the fields measured in the preferred frame and a non-preferred frame is further suppressed by a factor of $\beta^2$ in the non-relativistic limit. 

There are two points to be made about the above derivation. First of all, 
Eq.~\eqref{EFieldAxis} assumes that the charge has been at rest for 
sufficiently long so that our expression for the electrostatic field applies. The analysis of a moving charge would require taking into account the self-interactions with radiated photons that circle around the compact dimension and hit the charge back. Secondly, we have completely neglected cosmic expansion and approximated our universe as static. Modeling the paradox on an expanding cylinder (or any compact Friedmann-Robertson-Walker universe) introduces many subtleties, \cite{JannaMaulik,MovingWalls}.

\section{Acknowledgements}
The  authors would like to thank Justin Khoury for his supervision during this work.  We would also like to thank Allan Blaer, Brian Greene, Dan Kabat, Janna Levin, Maulik Parikh, and Amanda Weltman for helpful discussions.  This work was supported by the VIGRE program of the Columbia University Departments of Mathematics and Physics.

\end{document}